\documentstyle[prl,twocolumn,aps]{revtex}
\input{epsf}
\begin{document}
\draft
\twocolumn[\hsize\textwidth\columnwidth\hsize\csname @twocolumnfalse\endcsname
\title{Low energy transition in spectral statistics of 2D interacting
fermions}

\author{Pil Hun Song$^{(a)}$ and Dima L. Shepelyansky$^{(b)}$}

\address {$^{(a)}$Max-Planck-Institut f\"{u}r Kernphysik, Postfach~103980,
69029 Heidelberg, Germany\\
$^{(b)}$Laboratoire de Physique Quantique, UMR 5626 du CNRS, 
Universit\'e Paul Sabatier, F-31062 Toulouse Cedex 4, France}

\date{April 16, 1999; revised September 7, 1999}

\maketitle

\begin{abstract}
We study the level spacing statistics $P(s)$ 
and eigenstate properties of spinless fermions with
Coulomb interaction on a two dimensional lattice at constant
filling factor and various disorder strength.  In the limit of large 
lattice size, $P(s)$ undergoes a transition from the 
Poisson to the Wigner-Dyson distribution at a critical total energy
independent of the number of fermions.  This implies the emergence
of quantum ergodicity induced by interaction 
and delocalization in the Hilbert space at zero temperature.
\end{abstract}
\vskip1pc]

\narrowtext


The experimental observation of the metal-insulator transition 
in two dimensions (2D) by Kravchenko {\it et al.}\cite{krav94} has
attracted a great interest to interacting fermions in a
disordered potential.  Indeed, according to the well-established 
result\cite{and79}, all states are localized for noninteracting 
particles in 2D.  Therefore, in the view of the
experimental result\cite{krav94}, a new theory should be developed to
understand the interaction effects between the localized fermionic
states.  However, in spite of various theoretical attempts, a coherent theory
for such systems is still not available.  While for highly excited
states, it has been shown that the repulsive/attractive interaction 
can induce a delocalization of two interacting 
particles\cite{ds94,imry,pichard,oppen}, the properties of low energy 
states are not understood yet.  Recently, in addition to experimental and
theoretical investigations, a number of attempts have been made to
study these many fermionic systems through numerical 
simulations\cite{berk,vojta,benen}.  Even though several interesting
features have been reported, the systems studied there are not large
enough to observe interaction induced delocalization.

In this paper we use another numerical approach based on the analysis
of spectral properties of multi-particle fermionic systems.  Indeed,
Shklovskii {\it et al.} have shown that the level spacing statistics 
is a powerful tool to analyze the Anderson transition in disordered
systems\cite{shklov}.  When the states are localized, the levels are
not correlated and the statistics is given by the Poisson
distribution, $P(s) = P_P(s) = \exp(-s)$, while in the metallic
phase, the states are ergodic and the statistics is close to the
Wigner surmise, $P_W(s) = (\pi s/2)\exp(-\pi s^2/4)$.  The critical 
transition point is characterized by an intermediate statistics which
depends on the boundary conditions\cite{braun} and the spatial 
dimension of the system\cite{zhar}.  This approach has also been used to
determine the quantum chaos border and the interaction induced
thermalization in finite fermionic systems \cite{jacq} and 
to detect 
Anderson transition for two electrons with the Coulomb interaction on 
2D disordered lattice\cite{ds99,cuevas}.  All these
results demonstrate that the approach developed in \cite{shklov}
allows to investigate efficiently the transition from nonergodic (localized) 
to ergodic eigenstates.  

Here we use the above method to study the change of the spectral
statistics, $P(s)$, with excitation energy $E$ in a model of spinless
fermions with Coulomb interaction on 2D disordered lattice.  
The Hamiltonian reads
\begin{equation}
H = V\sum_{<ij>} a^\dagger_i a_j + \sum_i w_i n_i
+ U \sum_{i>j} \frac{n_i n_j}{r_{ij}},
\end{equation}
where $a^\dagger_i (a_i)$ is the fermion creation (annihilation) operator
at site $i$, $V$ is the hopping between the nearest neighbors,
the diagonal energies $w_i$ are randomly distributed
within the interval $[-W/2,W/2]$, $n_i=a^\dagger_i a_i$ is the
occupation number at site $i$ and $U$ is the strength of the
Coulomb interaction with $r_{ij}$ the interparticle distance.
The particles move in a 2D cell of size $L \times L$ with the
periodic boundary conditions applied.  The Coulomb interaction is taken 
between electrons in one cell of size $L$ and with 8 charge images
in nearby 8 cells as in \cite{ds99}.  The number of particles $N_p$
and the cell size were varied within the intervals 2 $\leq N_p \leq$ 20 
and $8 \leq L \leq 28$.  With the notations in the Eq.~(1), the
parameter $r_s$, which measures the ratio of the Coulomb energy to the
Fermi energy, $\epsilon_F$, is given by $r_s = U/(2V\sqrt{\pi\nu})$, 
where $\nu = N_p/L^2$ is the filling factor and $\epsilon_F = 4\pi \nu V$.  
The majority of our data have been 
obtained for $\nu \approx 1/32$ (nearest rational value) and for 
$U/V =2$, which corresponds to $r_s = 3.22$.  

To study the level spacing statistics $P(s)$, we generalize the 
approach used in \cite{ds99} for many particles.  First, 
one particle eigenstates (orbitals) at $U=0$ are obtained and the Hamiltonian
(1) is rewritten in this basis using the two-body interaction matrix
elements.  We consider the first $M$ orbitals from the lowest energy and
the Hamiltonian multi-particle matrix, constructed only from these
orbitals, is diagonalized at the final stage.  To the maximum, $N_p = 20$ 
particles with up to $M=42$ orbitals has been considered and the resulting 
matrix size is $N_m \approx 5000$.  This size is significantly smaller 
than $_M C_{N_p}$ since the condition $\sum_{i=1}^{N_p} m_i \leq
\sum_{i=1}^{N_p-1} i + M$ applies for the one-particle orbital index
$m_i$.  Such a pyramid rule allows one to use efficiently only low
energy multi-particle states and to make a striking reduction of the
resulting total matrix size without any serious modification of
the low energy states.  We checked that our level statistics results
at low energy are not sensitive to the variation of $M$ and $N_m$
(see for example insert in Fig.1).  To 
compute the spectral statistics $P(s)$ at a given total excitation energy 
$E$, which is counted from the ground state energy, disorder average
has been performed over $N_D = 5000$ (for low energy) and $N_D =
1000$ (for higher energy) configurations.  In this way, the total
statistics for $P(s)$ for small energy interval varies from $N_S =
10^4$ at low $E$ to $N_S = 3\times 10^5$ at high $E$.  
\vskip -0.5cm
\begin{figure}
\epsfxsize=3.6in
\epsfysize=3.0in
\epsffile{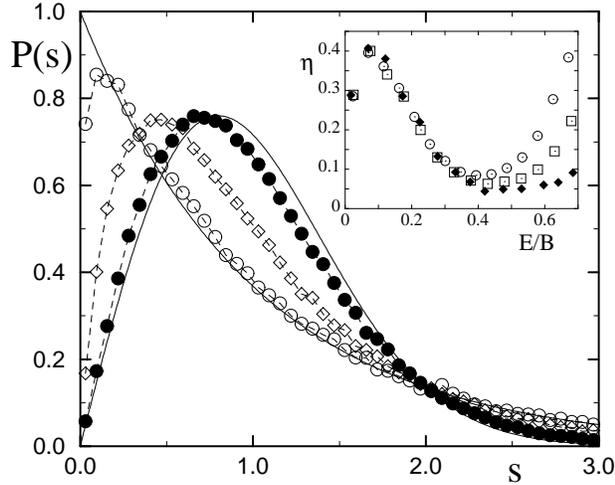}
\vglue -0.3cm
\caption{Level statistics $P(s)$ for
10 particles at $L=18$, $r_s=3.22$ in the
energy interval $0.25 <  E/B < 0.3$ (with $B=4V$):
$W/V=15$, $\eta=0.93$ ($\circ$); $W/V=10$, $\eta=0.51 \ (\diamond)$;
$W/V =7,\ \eta=0.10 \ (\bullet)$. Total statistics is
$N_S=10^5$; $1.4 \times 10^5$; $1.8 \times 10^5$,
respectively.  Full lines show the Poisson distribution
and the Wigner surmise. Insert, made  for $N_p=6$ case of Fig.2b,
illustrates that at low $E/B$ the statistics $P(s)$, 
characterized by $\eta$,
is independent of matrix size variation
$N_m =797 (\circ), 1231$ ($\Box$), 2235 (full diamond).
}
\label{fig1}
\end{figure}

A change of $P(s)$ with $W$, at a given $E$ and 
all other parameters fixed, is shown in Fig.~1 for $N_p =10$.  As 
$W$ decreases, $P(s)$ evolves from the Poissonian to the Wigner-Dyson
distribution.  To measure the proximity of $P(s)$ to one of these two
limiting cases, it is convenient to use the parameter $\eta$ which is
defined as $\eta=\int_0^{s_0}
(P(s)-P_{W}(s)) ds / \int_0^{s_0} (P_{P}(s)-P_{W}(s)) ds$,
where  $s_0=0.4729...$ is the smaller intersection point of $P_P(s)$ and
$P_{W}(s)$.  In this way $\eta=1$ corresponds to $P_P(s)$ 
and $\eta$=0 to $P_{W}(s)$.  According to Fig.~1,
$\eta$ changes almost by an order of magnitude when $W$ decreases by
factor of two.  This shows that for strong disorder, $W/V=15$, the
multi-particle states at given $E$ are not ergodic (localized) while for weaker
disorder, $W/V=7$, they become ergodic and characterized by the random
matrix spectral statistics.  We note that for given values of disorder
the one-particle inverse participation ratio (IPR), $\xi_1$ (the number of
sites contributing a one-particle eigenstate) is much smaller
than the total number of sites $L^2$ for the case of Fig.~1: 
\vskip -1.1cm
\begin{figure}
\epsfxsize=4.6in
\epsfysize=6.5in
\hskip -2.0cm
\epsffile{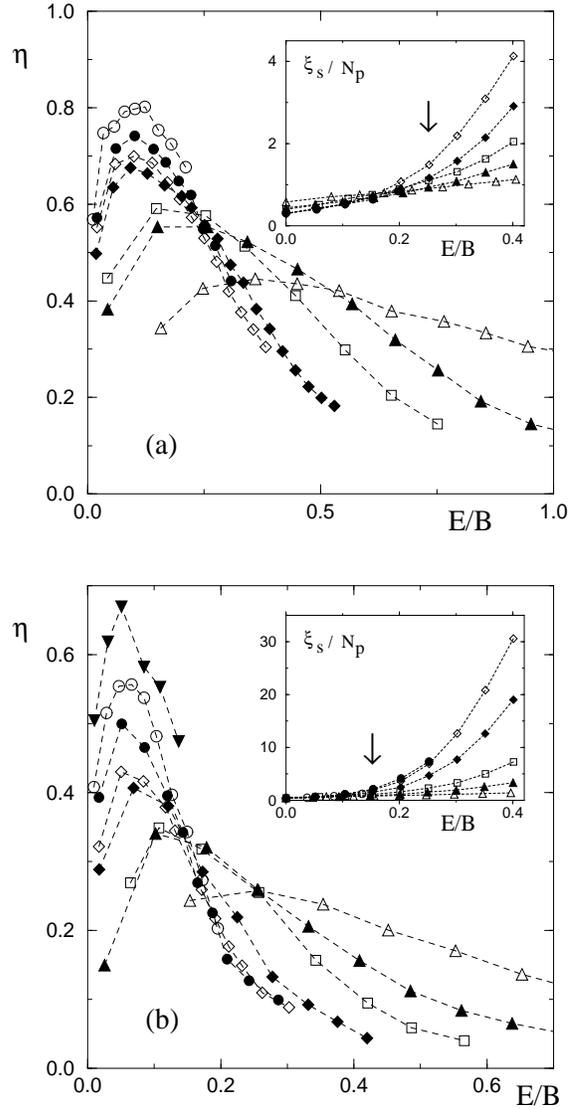}
\vglue -0.7cm
\caption{Dependence of $\eta$ on the rescaled total energy $E/B$ for
various number of particles $N_p = 2 (\bigtriangleup)$, 3 (full
triangle up), 4 ($\Box$),
6 (full diamond), 8 ($\diamond$), 10 ($\bullet$), 14 ($\circ$) and
20 (full triangle down); (a) $W/V = 10$ and (b) $W/V = 7$; filling
factor $\nu \approx 1/32$ and $r_s = 3.22$, $8\leq L\leq 25$.
Inserts show the dependence of $\xi_S/N_p$ 
on $E/B$  by same symbols; 
arrows mark the critical energy $E_c$ from the main figure.}
\label{fig2}
\end{figure}
$\xi_1$ = 3.4
and 4.2 ($W/V=15$); 5.2 and 11.6 ($W/V=10$); 8.2 and 36.7 ($W/V=7$), where
the formers are for the ground state and the latters for the center of
the band.  This means that the transition to ergodicity is induced by
the interaction which becomes effectively more strong when the
one-particle localization length increases (similar to the two
interacting particle case\cite{ds94,imry,pichard,oppen}).  At the same
time, the comparison with the data for $N_p=2$\cite{ds99} at the same
$E$ ($\eta \approx$ 0.92, 0.82 and 0.51 for $W/V$ = 15, 10 and 7, 
respectively) shows that the delocalization effect due to multi-particle
interaction is stronger for weak disorder while for strong disorder 
it does not affect localization.  
\begin{figure}
\epsfxsize=3.2in
\epsfysize=3.2in
\epsffile{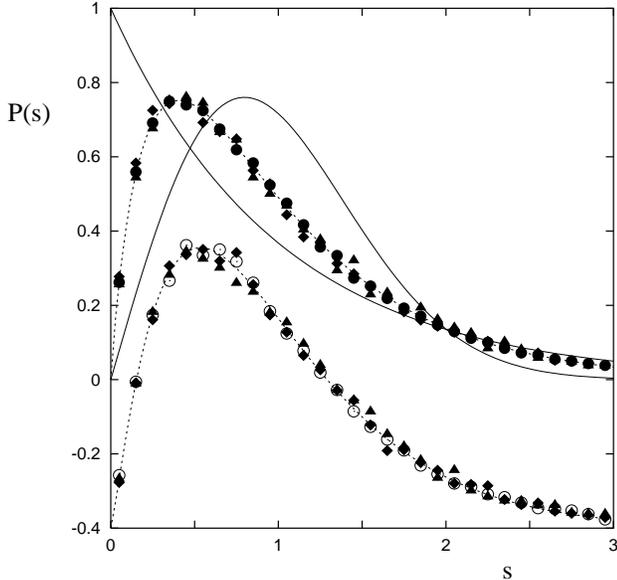}
\vglue -0.2cm
\caption{Level statistics $P(s)$ at the crossing points of Fig.~2
(with the same symbols).  The upper data are for $W/V = 10$, $ 0.225 < 
E/B < 0.275$, $\eta_c \approx 0.56$, $N_S =10^4, 4 \times 10^4$ and
$10^5$ for $N_p$ = 3, 6 and 10, respectively.  The lower data (shifted
down by 0.4) are for $W/V = 7$, $ 0.125 < E/B < 0.175$, $\eta_c \approx 
0.33$, $N_S = 10^4, 1.4 \times 10^4$ and $5 \times 10^4$ for 
$N_p$ = 3, 6 and 14, respectively.  The full lines are the Poisson 
distribution and the Wigner
surmise.}
\label{fig3}
\end{figure}

In Fig.~2 we present the variation of $\eta$ with the total energy $E$
for various number of particles at fixed filling factor and two values 
of disorder.  For $N_p > 2$, all curves have approximately 
the same crossing point: $E_c \approx 0.25 B \ (W/V=10)$, 
$E_c \approx 0.15 B \ (W/V=7)$
and $E_c \approx 0.1 B \ (\eta_c \approx 0.19, W/V=5$,  not shown).  
As $N_p$ increases, the statistics approaches the Poissonian case for $E < E_c$,
while it goes to the Wigner case for $E > E_c$.  The transition
point, $E_c$, is characterized by an intermediate statistics, which is
independent of the system size as shown in Fig.~3.  The energy $E_c$ 
grows with the increase of $W$ so that for $W/V=15$ the crossing point 
is not found clearly within our computations (in this case $\eta
\approx$ 0.8 - 1.0 for $0 < E/B \leq 1$ and $N_p > 2$).  The above data give a
strong evidence that the ground state and the excited states with $E <
E_c$ are localized in the limit of the infinite system size. 
However, for excitations with $E > E_c$ the statistics is close to
$P_W(s)$ that implies the ergodicity of eigenstates and their
space delocalization.  

To show that the eigenstate properties 
are qualitatively different below and above $E_c$ we determined 
the variation of IPR $\xi_S$ with $N_p$ and $E$ as presented 
in the inserts of Fig.2. This IPR $\xi_S$ is
computed in the basis of noninteracting Slater determinants
and determines how many of such states are required to construct
an eigenstate. The results are obtained by averaging
over 100 disorder realizations so that the statistical
error is less than $ 5\% $.
For $E<E_c$ the ratio $\xi_S/N_p$, which gives the
number of noninteracting states per particle, remains
constant while for $E>E_c$ it grows rapidly with $N_p$.
For example, the ground state  average value is 
$\xi_S/N_p=0.35; 0.40; 0.46$
for $W/V=10; 7; 5$ (the latter is not shown) with only
$\pm 21; 13; 18 \%$ variation when $N_p$ changes from 
3 to 10; 14; 20, respectively for each $W/V$. On the contrary,
for $E=0.4 B > E_c$ the ratio  $\xi_S/N_p$ grows in 
2.7; 9; 20  times when $N_p$ changes from 3 to 8
for $W/V=10; 7; 5$ respectively. These results 
show that the delocalization in the
Hilbert space of Slater determinants
takes place for $E>E_c$ with $E_c$ independent
of number of particles. A crossover 
from localized to extended states in the Hilbert space
was extensively discussed recently for the 
metallic phase with extended one-particle states \cite{hil,jacq}.
Our data in Fig. 2 represent the first evidence for a 
delocalization transition in the Hilbert space
for the insulating phase with localized one-particle states.
\vskip -1.5cm
\begin{figure}
\epsfxsize=3.2in
\epsfysize=4.55in
\epsffile{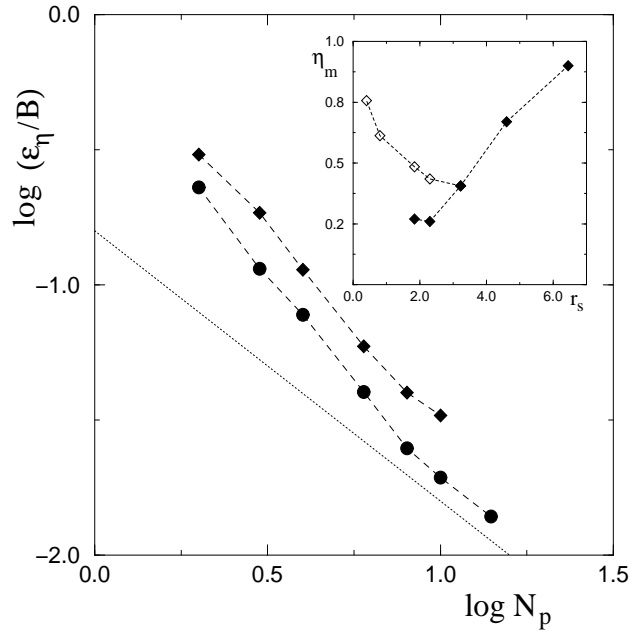}
\vglue -1.5cm
\caption{Dependence of $\epsilon_\eta/B$ on the number of 
particles $N_p$, obtained from Fig.~2: 
$W/V = 10$ with $\eta \ (E_\eta) = 0.4$ (full diamond) and $W/V
= 7$ with $\eta \ (E_\eta) = 0.2$ ($\bullet$), where $\epsilon_\eta = 
E_\eta/N_p$.  The straight line shows the slope when $E_\eta = const.$ 
Insert gives the dependence of maximal $\eta$ on $r_s$ for $W/V =7$
and $N_p=6$: $U/V =2$, $8 \leq L \leq 28$ (full diamond) and 
$L=14,\ 0.25 \leq U/V \leq 2$ $(\Diamond)$.  
}
\label{fig4}
\end{figure}

An unusual property of the above transition is that it takes place at the
finite total energy.  This means that the energy per particle
$\epsilon = E/N_p$, or the temperature, is equal to zero at the transition
point as $N_p \rightarrow \infty$.  
In this sense this transition can be considered as a quantum zero
temperature transition.  To ensure that in our numerical computations
$\epsilon$ was sufficiently low, we present variation of $\epsilon_\eta$, 
defined at the fixed $\eta$ level, with the number of particles in Fig.~4.
According to these data $\epsilon_\eta$ drops more than by an order of
magnitude and becomes approximately by an order of magnitude smaller 
than the Fermi energy $\epsilon_F \approx 0.1 B$.  

Dependence of $\eta_m$ on $r_s$ is found in the insert of Fig.~4.
The value of $r_s$ is varied in two different ways: i) by changing the
cell size $L$ or ii) by changing $U/V$.  The first case i) allows to obtain
large $r_s$ value and shows that the statistics approaches the
Poissonian case.  This result is in a qualitative agreement with the
experimental data\cite{krav94}, where the localized (insulating) phase
appears for large $r_s$, and with the theoretical argument given in
\cite{ds99}, according to which the two-body interaction drops as
the filling factor decreases.  The same arguments explain
why $\eta$ goes to 1 when the interaction strength $U$ decreases as in the
second case ii).  One can expect that the variation of $\eta_c$ with $r_s$
at large $N_p$ and fixed $\nu$ will be qualitatively similar
to that one of $\eta_m$ in Fig.~4.

The comparison of the energy $E_c$, at which the transition at fixed
$\nu$ takes place (see Fig.~2), with the transition energy 
for two particles ($ E_{2c}/B=1.2$ for $W/V$=10 and $ E_{2c}/B=0.56$
for $W/V$=7 \cite{ds99}) shows that $E_c$ is significantly smaller
than $E_{2c}$ ($E_c/E_{2c} \approx 0.2$).  This signifies that the
multi-particle interaction is more efficient for delocalization than
the two-body interaction for only two particles.  Our qualitative
understanding for the transition to the quantum ergodicity at fixed
$\nu$ and fixed total energy $E_c$ is based on the following scenario.
In the analogy with the approach developed in \cite{ds99} for $N_p=2$,
the dynamics of three or four particles at finite energy $E$ can be
assumed to be equivalent to one particle dynamics in a disordered 
system with effective spatial dimension $3 < d_{eff} < 2 N_p$.  With 
the growth
of $E$, the rate of this one-particle hopping increases\cite{ds99} and
the delocalization transition finally takes place.  As a result, in
the original system, the energy can be transferred by this small group of
particles from one place to another that allows to
establish the ergodicity in the system of large size at fixed filling
factor.  The values of $\eta_c$ found here (Fig.~2) are larger than that
for 3D Anderson model ($\eta_c \approx 0.2$), that makes the above scenario 
consistent with the result of \cite{zhar} according to which $\eta_c$ 
grows with the increase of the spatial dimension $d$.  
We also note that the approach of $\eta$ to 1 near the
ground state is not so surprising.  Indeed, even in the metallic quantum
dots one has   $\eta \approx 1$ near the ground state 
since the interaction between particles is
effectively weak\cite{jacq,hil} and the transition to ergodicity 
with $\eta = 0$ takes place only at higher energy\cite{note1}.

In conclusion, our results show that 2D fermions, which are strongly 
localized by disorder without interaction, become ergodic due to
Coulomb interaction.  The transition to ergodicity, 
in lattice and Hilbert spaces, and the random
matrix statistics takes place at the finite total energy or zero
temperature.  These results are in favor of the experimentally
observed metal-insulator transition\cite{krav94}, even if our approach
does not allow to investigate the conductance dependence on
temperature which is the main experimental method to detect the 
transition. Furthermore, they show that the variable 
range hopping transport can be induced by electron-electron interaction
in agreement with recent experiments \cite{shlimak}.

We acknowledge the IDRIS at Orsay for allocation of the CPU time on
the supercomputers.  We also thank the Max-Planck-Institut f\"{u}r
Physik Komplexer Systeme at Dresden for the hospitality at the final 
stage of this work during the workshop Dynamics of
Complex Systems.


\begin{thebibliography}{99}
\bibitem{note1} We can try to adopt the approach\cite{jacq} 
        to the present case.
        According to\cite{ds99} the two-body matrix element between
        particles at a distance $L$ is $U_s \sim U/L^3$ and the number
        of such effectively interacting particles 
        at temperature $T$ is $n_{eff} \sim \nu L^2 T/\epsilon_F$.
        Then the spacing between directly coupled multi-particle states is
        $\Delta_c \sim \Delta_2/n_{eff}^2$ where at low energy 
        the two-particle spacing is $\Delta_2 \sim V/\xi_1$,
        since each particle can jump only inside $\xi_1$ 
        sites \cite{imry,ds99}.
        The transition to $\eta =0$ takes place for $U_s > \Delta_c$\cite{jacq}
        that gives the critical energy 
        $E_c \sim T n_{eff} \sim (N_p/\nu)^{1/2} V^2/U\xi_1$.
        However, this border is higher than the numerically found one.
        This indicates that the delocalization of small
        group of particles is dominating.
\end{thebibliography}
\end{document}